\documentclass[12pt]{article}

\usepackage{amssymb,amsmath,graphics}

\numberwithin{equation}{section}

\allowdisplaybreaks

\newenvironment{Proof}{\removelastskip\par\medskip
\noindent{\em Proof.}
\rm}{\penalty-20\null\hfill$\square$\par\medbreak}

\newtheorem{prop}{Proposition}
\newtheorem{lemma}{Lemma}

\newtheorem{corollary}{Corollary}

\newtheorem{remark}{Remark}

\def\max{{\mathord{\mathrm {Max}}}}

\begin{document}
  ~
\begin{center}
{\Large On option pricing in illiquid markets with jumps}
\\
\bigskip
{\large Youssef El-Khatib}\footnote{UAE University, Department of Mathematical Sciences, Al-Ain, P.O. Box 17551. United Arab Emirates.{\ E-mail : Youssef\_Elkhatib@uaeu.ac.ae.}
}\ \ \ \ \ {\large Abdulnasser Hatemi-J}\footnote{UAE University, Department of Economics and Finance, Al-Ain, P.O. Box 17555, United Arab Emirates. {\ E-mail : Ahatemi@uaeu.ac.ae.}}
\end{center}
~
\begin{abstract}
One of the shortcomings of the Black and Scholes model on option pricing is the assumption that trading of the underlying asset does not affect the price of that asset. This assumption can be fulfilled only in perfectly liquid markets. Since most markets are illquid, this assumption might be too restrictive. Thus, taking into account the price impact in option pricing is an important issue. This issue has been dealt with, to some extent, for illiquid markets by assuming a continuous process, mainly based on the Brownian motion. However, the recent financial crisis and its effects on the global stock markets have propagated the urgent need for more realistic models where the stochastic process describing the price trajectories involves random jumps. Nonetheless, works related to markets with jumps are scant compared to the continuous ones. In addition, these previous studies do not deal with illiquid markets. The contribution of this paper is to tackle the pricing problem for options in illiquid markets with jumps as well as the hedging strategy within this context, which is the first of its kind to the best knowledge.
\end{abstract}

\baselineskip=0.5cm

\baselineskip=0.5cm \noindent\textbf{Keywords:} Options pricing, illiquid markets, jump diffusion, incomplete markets.
\baselineskip0.7cm

\section{Introduction}

\label{sec1} Financial derivatives are important tools for dealing with financial risk. An option is an example of such derivatives, which gives the right but not the obligation, to engage in a future transaction on some underlying financial asset. For instance, a European call option on an asset with the price $(S_t)_{t\in[0,T]}$-
is a contract between two agents (buyer and seller), which gives the holder the right to buy the asset at a pre-specified future time $T$ (the expiration date) for an amount K (called the strike). The buyer of the option is not obliged to exercise the option. When the contract is issued they buyer of the option needs to pay a certain amount of money called the premium. The payoff for this option is defined as $h(S_T)=\max(S_T-K,0)=(S_T-K)^+$. The writer of the option receives a premium that is invested in the combination of the risky and risk free assets. The pricing problem is then to determine the premium, i.e. the price that the seller should charge for this option. \\
The pricing problem has been solved in the pioneer work of Black and Scholes\cite{BS}. One of the shortcoming of the Black and
Scholes model is the assumption that an option trader cannot affect the
underlying asset price. However, it is well-known that in a market with
imperfect liquidity, trading does affect the underlying asset price (see,
for example, Chan and Lakonishok\cite{ChaLak}, Keim and Madhavan\cite{KeiMad}, and Sharpe et al.\cite{Shal}). \newline
In Liu and Yong\cite{Liuy}, the authors study the effect of the replication of a
European option on the underlying asset price.  They obtain a generalization of the Black Scholes pricing P.D.E. as the following:
\begin{eqnarray}
&&\frac{\partial v}{\partial t}(S,t)+\frac{\sigma ^{2}S^{2}}{2\left(
1-\lambda (S,t)S\frac{\partial ^{2}v}{\partial S^{2}}(S,t)\right) ^{2}}\frac{%
\partial ^{2}v}{\partial S^{2}}(S,t)+r\frac{\partial v}{\partial S}%
(S,t) \nonumber \\
&&-r_{t}v(S,t)=0, \ \ \ \ \ \ \ \ \ \ \mbox{for }(S,t)\in ]0,+\infty \lbrack \times ]0,T]  \label{1} \\
&&v(S,T)=f(S),\ \ \ \ \ \ \ \ \ \ 0<S<\infty,  \label{2}
\end{eqnarray}%
where $\lambda (S,t)$ is the price impact function of the trader. The classical Black–Scholes P.D.E. is a special case of (\ref{2}) when $\lambda (S,t)=0$.\\

There are also several other papers that have studied the financial markets with jumps (among others, Merton\cite{Merton}, Dritschel and Protter\cite{DP}, El-Khatib and Privault\cite{elkhatibpriv}) and El-Khatib and Al-Mdallal\cite{khatib1}. However, none of the previous studies based on the jump-diffusion approach deals with illiquid markets, to the best knowledge. This paper is extends the model of Liu and Yong\cite{Liuy} by including a jump-diffusion structure in the underlying option pricing model. This appears to be an important issue because the model that is suggested in this paper allows for the possibility to account for sudden and random significant changes in the market that might not be captured by the existing models in the literature such as the continuous model suggested by Liu and Yong\cite{Liuy}. Hence, the approach that is developed in this paper is expected to be more useful in financial risk management, especially in the cases in which the financial markets are under stress.\\

The disposition of the rest of the paper is the following. Section~2 introduces the jump-diffusion model for an illiquid market. Section~3 deals with the pricing problem of an option within the context of a jump-diffusion model along with the proof for the suggested solution. Section 4 concludes the paper.

\section{A jump-diffusion model for illiquid markets}

\label{s1}
We start with presenting some necessary denotations. Let $(N_t)_{t \in [0,T]}$ be a Poisson process with deterministic intensity $\rho$. Let also $M_t=N_t - \rho t$ be its associated compensated process. The process $(B_t)_{t\in [0,T]}$ denotes a Brownian motion. The probability space of interest is $(\Omega ,{\cal F}, P)$ with $(M_t)_{t \in [0,T]}$ and $(B_t)_{t \in [0,T]}$  being independent. Let $({\cal F}_t)_{t \in [0,T]}$ signify the filtration generated by $(N_t)_{t \in [0,T]}$ and $(B_t)_{t\in [0,T]}$. The market is assumed to have two assets: a risky asset $(S_t)_{t \in [0,T]}$ and a risk-free denoted by $(A_t)_{t \in [0,T]}$. The maturity is $T$, the strike is $K$ and the payoff is $h(S_{T})=(S_{T}-K)^{+}\equiv \max\{S_{T}-K,0\}$. As in Liu and Yong\cite{Liuy}, the return on the risk free asset indirectly depends on $S_t$ and the option trader's trading in the stock market has a direct impact on the stock price. This price impact, which an investor can cause by trading on an asset, functions in such way that it increases the price when buying the asset and it decreases the price when selling the asset. The price of the risk-free asset is given by
\begin{equation}
\label{stock0}
dA_{t}=r(t,S_t)A_{t}dt,\ \ \ t\in [0,T],
\end{equation}
where $r>0$ denotes the interest rate. The price of the risky asset is generated by the following stochastic differential equation:
\begin{equation}
\label{stock}
dS_{t}=\mu(t,S_{t})dt+\sigma(t,S_{t})(dW_{t}+ a dM_{t})+\lambda(t,S_t)d\theta_t,\ \ \
t\in [0,T],\ \ \ S_{0}=x>0,
\end{equation}
where $\mu$ and $\sigma$ represent the expected return and volatility, respectively, the term $a$ is a real constant and $\lambda(S,t)$ denotes the price impact factor created by the trader via selling or buying the underlying asset. $\theta_t$ is the number of shares that the trader has in the stock at time $t$. Hence, $\lambda (S,t)d\theta_t$, captures the price impact of trading. Before dealing with the pricing of a European option in a jump-diffusion illiquid market, we need to observe the following remark.
\begin{remark}
The parameter $a$ in (\ref{stock}) determines the direction of the jumps\footnote{it affects also the jumps size.}. In fact the following can be stated:
\begin{itemize}
\item If $a<0$ the jumps are pushing the stock price down, i.e. the stock price is decreasing at each jump.
\item If $a =0$ then there are no jumps and therefore model (\ref{stock}) is reduced to the model in Liu and Yong\cite{Liuy}.
\item If $a>0$ the jumps are pushing the stock, i.e. the stock price is increasing at each jump
\end{itemize}
\end{remark}

\section{Pricing of a European option in jump-diffusion illiquid market}

Let $(V_t)_{t \in [0,T]}$ be the wealth process for the trader. Let also $(\psi_t)_{t \in [0,T]}$ denote the number of shares invested in the risk-free asset. Then, the value of the portfolio is given by
\begin{equation}
\label{wealth}
V_t=\psi_t A_t+\theta_t S_t,\ \ \  t \in [0,T].
\end{equation}
Assume that the number of shares of the risky asset satisfies the following condition:
\begin{equation}
\label{numbshare}
d\theta_t=\eta_t dt +\zeta_t (dW_t+b dM_t), \  \  \ t \in [0,T].
\end{equation}
Let us consider a European call option with the payoff defined as $h(S_T):=(S_T-K)^+$. In order to replicate the option for a perfect hedge, we search for a strategy $(\psi_t,\theta_t)_{t \in [0,T]}$ which, at the expiration date of the option, leads to having a value of the underlying wealth to be equal to the payoff, that is  $V_T=h(S_T)$. Then we can state the following corollary.
\begin{corollary}
The wealth process for the trader of the jump-diffusion model in section~\ref{s1} satisfies the following stochastic differential equation:
\begin{eqnarray}
\nonumber
dV_t&=&\left\{r(t,S_t)V_t +\left[\mu(t,S_t)-r(t,S_t)+\lambda(t,S_t)\eta_t\right]\theta_t S_t\right\}dt\\
\label{wealth1}
   &&+\theta_t S_t[\lambda(t,S_t)\zeta_t+\sigma(t,S_{t})]dW_t+\theta_t S_t[a\sigma(t,S_{t})+b\lambda(t,S_t)\zeta_t]dM_t
\end{eqnarray}
\end{corollary}
\begin{Proof}
By using equations (\ref{stock0}), (\ref{stock}), (\ref{wealth}) and (\ref{numbshare}) we have the following:
\begin{eqnarray*}
\nonumber
dV_t&=&\psi_t dA_t+\theta_t dS_t\\
\nonumber
   &=&\frac{V_t-\theta_t S_t}{A_t} dA_t+\theta_t S_t \left[\mu(t,S_{t})dt+\sigma(t,S_{t})(dW_{t}+ a dM_{t})+\lambda(t,S_t)d\theta_t\right]\\
   &=&\left\{r(t,S_t)V_t +\left(\mu(t,S_t)-r(t,S_t)\right)\theta_t S_t\right\}dt+\theta_t S_t\left\{\sigma(t,S_{t})(dW_{t}+ a dM_{t})\right.\\
   &&+\left.\lambda(t,S_t)\left[\eta_t dt +\zeta_t (dW_t+b dM_t)\right]\right\}\\
   &=&\left\{r(t,S_t)V_t +\left[\mu(t,S_t)-r(t,S_t)+\lambda(t,S_t)\eta_t\right]\theta_t S_t\right\}dt\\
   &&+\theta_t S_t[\lambda(t,S_t)\zeta_t+\sigma(t,S_{t})]dW_t+\theta_t S_t[a\sigma(t,S_{t})+b\lambda(t,S_t)\zeta_t]dM_t,
\end{eqnarray*}
which ends the proof.
\end{Proof}
Our aim in this paper is to price the European option with payoff $h(S_T)$ where $S_T$ is given by (\ref{stock}). We replicate the European option by searching a wealth $(V_t)_{t \in [0,T]}$ which leads to the terminal value $V_T=h(S_T)$. Thus, as in Liu and Yong\cite{Liuy}, we need to solve the following system of stochastic differential equations.
\begin{eqnarray}
\nonumber
d\theta_t &=&\eta_t dt +\zeta_t (dW_t+b dM_t),\\
\nonumber
\frac{dS_{t}}{S_t}&=&[\mu(t,S_{t})+\lambda(t,S_t)\eta_t]dt+[\sigma(t,S_{t})+\lambda(t,S_t)\zeta_t]dW_{t}\\
\nonumber
&&+ [a\sigma(t,S_{t})+b\lambda(t,S_t)\zeta_t]dM_{t},\\
\nonumber
dV_t&=&\left\{r(t,S_t)V_t +\left[\mu(t,S_t)-r(t,S_t)+\lambda(t,S_t)\eta_t\right]\theta_t S_t\right\}dt\\
\nonumber
   &&+\theta_t S_t[\lambda(t,S_t)\zeta_t+\sigma(t,S_{t})]dW_t+\theta_t S_t[a\sigma(t,S_{t})+b\lambda(t,S_t)\zeta_t]dM_t,\\
\label{FB}
\theta_0 &>&0, \ \ \ S_0>0, \ \ \ V_T=h(S_T),
\end{eqnarray}
The above system is called FBSDE (forward-backward stochastic differential equations) system. In order to derive the P.D.E. for the European option price, we need  It\^o formula which is given by the following lemma (see Protter\cite{protter1}).
\begin{lemma}
Let $g$, $l$, and $k$ be three adapted processes such that
$$
\int_0^t |g_s|ds<\infty, \ \ \  \int_0^t |l_s|^2ds<\infty, \ \ \
\mbox{and} \ \ \ \int_0^t \rho |k_s|ds<\infty.
$$
Let $X=(X_t)_{t\in [0,T]}$ be the process defined by
$$
dX_t=g_t dt +l_t dW_t +k_t dM_t. \ \ \
$$
For any function $G\in \mathcal{C}^{1,2}([0,T]\times
]-\infty,\infty[$, we have
\begin{eqnarray}
G(t,X_t)&=&G(0,X_0)+\int_0^t\left(\partial_s G
(s,X_s)+(g_s-k_s\rho)\partial_x G(s, X_{s^-})\right.\nonumber\\
&&+\left.\frac{1}{2}l^2_s \partial^2_{xx} G(s, X_{s^-})\right)ds+\int_0^t l_s \partial_x G(s, X_{s^-})dW_s\nonumber\\
&&+\sum_{s\leq t}\left(G(s,X_s)-G(s,X_{s^-})\right).\label{ito}
\end{eqnarray}
\end{lemma}
Equation (\ref{ito})  can be written in the following format:
\begin{eqnarray}
G(t,X_t)&=&G(0, X_0)+\int_0^t\left[\partial_s G
(s,X_s)+(g_s-k_s\rho)\partial_x G(s,X_{s^-})
\nonumber
+\right.\\
\nonumber
&&\left.\frac{1}{2}l^2_s \partial^2_{xx} G(s,X_{s^-})+\rho(G(s,X_{s^-}+k_s)-G(s,X_{s^-}))\right]ds\\
\nonumber
&&+\int_0^t [G(s,X_{s^-}+k_s)-G(s,X_{s^-})]dM_s\\
\label{itom}
&&+\int_0^t l_s \partial_x G(s,X_{s^-})dW_s.
\end{eqnarray}
The next proposition provides the P.D.E. for the price of the European option in the jump-diffusion illiquid market presented in section.~\ref{s1}.
\begin{prop}
Let $f(t,S_t)$ denote the price of the European option at time $t \in [0,T]$ for the model presented in section.~\ref{s1}. Then the corresponding P.D.E. for the underlying option price is given by
\begin{eqnarray*}
\nonumber
\lefteqn{r(t,S_t)V_t +\left[\mu(t,S_t)-r(t,S_t)+\lambda(t,S_t)\eta_t\right]\theta_t S_t=}\\
&&\partial_t f
(t,S_t)+\left(\mu(t,S_{t})+\lambda(t,S_t)\eta_t-\rho[a\sigma(t,S_{t})+b\lambda(t,S_t)\zeta_t]\right)S_t\partial_S f(t,S_{t})\\
\nonumber
&&+\frac{1}{2}[\sigma(t,S_{t})+\lambda(t,S_t)\zeta_t]^2 S^2_{t}\partial^2_{SS} f(t,S_{t})
+\rho\left(f\left(t,S_{t^-}(1+a\sigma(t,S_{t})\right.\right.\\
&&\left.\left.+b\lambda(t,S_t)\zeta_t)\right)-f(t,S_{t^-})\right),
\end{eqnarray*}
with the terminal condition $f(T,S_T)=h(S_T).$
Moreover, the market is incomplete and there is no strategy leading to the terminal wealth $V_T=h(S_T):=f(T,S_T)$. However, the number of shares $\theta$ that minimizes the variance is given by
$$
\theta_t=\frac{(\sigma+\lambda\zeta)^2S^2\partial_S f + \rho S(a\sigma+b\lambda\zeta) \left(f\left(t,S_{t^-}(1+a\sigma+b\lambda\zeta)\right)-f\right)}{(\sigma+\lambda\zeta)^2S^2+\rho S^2(a\sigma+b\lambda\zeta)^2}.
$$
\end{prop}
\begin{Proof}
Let $(\theta, S, V)$ be an adapted solution of the FBSDE (\ref{FB}) and assume that there exists a smooth function $f\in {\mathcal{C}}^{3,1}(]-\infty,\infty[\times [0,T])$ such that $f(t,S_t)$ represents the price of the European option at time $t \in [0,T]$. Since the price of the option at maturity is equal to the payoff, then $f(T,S_T)=h(S_T)$. Now, using It\^o formula (\ref{itom}) we obtain
\begin{eqnarray}
\nonumber
df(t,S_t)&=&\left\{\left(\mu(t,S_{t})+\lambda(t,S_t)\eta_t-\rho[a\sigma(t,S_{t})+b\lambda(t,S_t)\zeta_t]\right)S_t\partial_S f(t,S_{t})\right.\\
\nonumber
&&+\left.\frac{1}{2}[\sigma(t,S_{t})+\lambda(t,S_t)\zeta_t]^2 S^2_{t}\partial^2_{SS} f(t,S_{t})+\partial_t f(t,S_t)\right.\\
\nonumber
&&+\left.\rho\left(f\left(t,S_{t^-}(1+a\sigma(t,S_{t})+b\lambda(t,S_t)\zeta_t)\right)-f(t,S_{t^-})\right)\right\}dt\\
\nonumber
&&+[\sigma(t,S_{t})+\lambda(t,S_t)\zeta_t]S_{t}\partial_S f(t,S_{t})dW_t\\
\label{wealth2}
&&+[f\left(t,S_{t^-}(1+a\sigma(t,S_{t})+b\lambda(t,S_t)\zeta_t)\right)-f(t,S_{t^-})]dM_t.
\end{eqnarray}
By comparing equations (\ref{wealth1}) and (\ref{wealth2}) one can deduce that it is impossible to find a strategy $(\eta_t,\zeta_t)_{t \in [0,T]}$ that results in the terminal wealth $V_T=h(S_T):=f(T,S_T)$. Thus, we put the term belonging to $dt$ equations (\ref{wealth1}) and (\ref{wealth2}) equal to each other, which gives the P.D.E. of the option price and then we minimize the distance between the wealth $V_T$ and the price $f(T,S_T)=h(S_T)$ over the number of shares of the underlying asset, i.e. $\theta_t$. The P.D.E. of the option price in this case is
\begin{eqnarray}
\nonumber
\lefteqn{r(t,S_t)V_t +\left[\mu(t,S_t)-r(t,S_t)+\lambda(t,S_t)\eta_t\right]\theta_t S_t=}\\
\nonumber
&&\partial_t f
(t,S_t)+\left(\mu(t,S_{t})+\lambda(t,S_t)\eta_t-\rho[a\sigma(t,S_{t})+b\lambda(t,S_t)\zeta_t]\right)S_t\partial_S f(t,S_{t})\\
\nonumber
&&+\frac{1}{2}[\sigma(t,S_{t})+\lambda(t,S_t)\zeta_t]^2 S^2_{t}\partial^2_{SS} f(t,S_{t})
+\rho\left(f\left(t,S_{t^-}(1+a\sigma(t,S_{t})\right.\right.\\
\label{pde0}
&&+\left.\left.b\lambda(t,S_t)\zeta_t)\right)-f(t,S_{t^-})\right),
\end{eqnarray}
with the terminal condition $$f(T,S_T)=h(S_T).$$
To find the number of shares $\theta_t$ invested in $S_t$ we need to solve the following problem:
\begin{equation}
\label{minim}
\mbox{Minimize}_{\theta} E[\Pi^2(\theta)],
\end{equation}
where $\Pi(\theta):=(h(S_T)-V_T)$.
By using (\ref{wealth1}), (\ref{wealth2}) and (\ref{pde0}) we have
\begin{eqnarray*}
 E[\Pi^2(\theta)]&=&E\left[\left(\int_0^T \left([\sigma(t,S_{t})+\lambda(t,S_t)\zeta_t]S_{t}(\partial_S f(t,S_{t})-\theta_t)\right)dW_t\right)^2\right]\\
&&+E\left[\left(\int_0^T \left(f\left(t,S_{t^-}(1+a\sigma(t,S_{t})+b\lambda(t,S_t)\zeta_t)\right)-f(t,S_{t^-})\right.\right.\right.\\
&&-\left.\left.\left.\theta_t S_t[a\sigma(t,S_{t})+b\lambda(t,S_t)\zeta_t]\right)dM_t\right)^2\right]\\
&=&E\left[\int_0^T \left([\sigma(t,S_{t})+\lambda(t,S_t)\zeta_t]S_{t}(\partial_S f(t,S_{t})-\theta_t)\right)^2dt\right]\\
&&+E\left[\int_0^T \rho\left(f\left(t,S_{t^-}(1+a\sigma(t,S_{t})+b\lambda(t,S_t)\zeta_t)\right)-f(t,S_{t^-})\right.\right.\\
&&\left.\left.-\theta_t S_t[a\sigma(t,S_{t})+b\lambda(t,S_t)\zeta_t]\right)^2 dt\right]\\
&=&E\left[\int_0^T l(\theta_t) dt \right],
\end{eqnarray*}
where
$$
l(x)= (\sigma+\lambda\zeta)^2S^2(\partial_S f-x)^2+\rho\left(f\left(t,S_{t^-}(1+a\sigma+b\lambda\zeta)\right)-f-xS(a\sigma+b\lambda\zeta)\right)^2.
$$
The minimum is obtained at $l^{'}(x)=0$, which yield the following result:
\begin{eqnarray*}
&&2(\sigma+\lambda\zeta)^2S^2(\partial_S f-x)-2 S(a\sigma+b\lambda\zeta) \rho\left(f\left(t,S_{t^-}(1+a\sigma+b\lambda\zeta)\right)-f\right.\\
&&\left.-x S[a\sigma+b\lambda\zeta]\right)=0,
\end{eqnarray*}
and
$$
\theta_t=\frac{(\sigma+\lambda\zeta)^2S^2\partial_S f + \rho S(a\sigma+b\lambda\zeta) \left(f\left(t,S_{t^-}(1+a\sigma+b\lambda\zeta)\right)-f\right)}{(\sigma+\lambda\zeta)^2S^2+\rho S^2(a\sigma+b\lambda\zeta)^2},
$$
which ends the proof.
\end{Proof}
It is worth mentioning that in the case where there are no jumps, i.e. when $a=b=0$, then $\theta =\partial_S f$ and the P.D.E. in the previous proposition is reduced to the P.D.E. that is obtained in Liu and Yong\cite{Liuy}, assuming there are no dividends.

\section{Conclusion}

Option pricing is an integral part of modern risk management in increasingly globalized financial markets. The classical Black and Scholes model is regularly used for this purpose. However, one of the main pillars that makes this model operational is the underlying assumption that the markets are perfectly liquid. This assumption is, nonetheless, not fulfilled in reality since perfectly liquid markets do not exist. In our opinion the question should not be whether the markets are illiquid or not, the question should be about the degree of illiquidity. Thus, taking into account the fact that markets are illiquid can improve on the precision of the underlying option pricing.

This paper is the first attempt, to our best knowledge, that extends the existing literature on option pricing by introducing a jump-diffusion model for illiquid markets. This seems to be a more realistic approach to deal with a market that is incomplete. A solution for the option pricing within this context is provided along with the underlying proof. The suggested solution might be useful to investors in order to determine the optimal value of an option in a market that is characterized by illiquidity.

\end{document}